\documentclass[11pt,a4paper]{article}

\usepackage[utf8]{inputenc}
\usepackage{graphicx}
\usepackage{amsmath, amssymb}
\usepackage{hyperref}
\usepackage{booktabs}
\usepackage{geometry}
\usepackage{titlesec}
\usepackage{authblk}
\usepackage{xcolor}
\usepackage{orcidlink}
\usepackage{float}

\titlespacing*{\section}{0pt}{1.2ex plus 0.5ex minus .2ex}{0.8ex plus .2ex}
\titlespacing*{\subsection}{0pt}{1.0ex plus 0.3ex minus .2ex}{0.6ex plus .2ex}

\title{\textbf{Enhancing Hotel Recommendations with AI: LLM-Based Review Summarization and Query-Driven Insights}}

\author[1]{Nikolaos Belibasakis\thanks{\texttt{belibasakis@ceid.upatras.gr}}\,\orcidlink{0009-0008-3895-0032}}
\author[1]{Anastasios Giannaros\thanks{Corresponding author: \texttt{giannaros@ceid.upatras.gr}}\,\orcidlink{0000-0001-9413-8841}}
\author[2]{Ioanna Giannoukou\thanks{\texttt{igiannoukou@upatras.gr}}\,\orcidlink{0000-0003-2112-6430}}
\author[1]{Spyros Sioutas\thanks{\texttt{sioutas@ceid.upatras.gr}}\,\orcidlink{0000-0003-1825-5565}}

\affil[1]{Computer Engineering and Informatics Department, University of Patras, Greece}
\affil[2]{Department of Management Science and Technology, University of Patras, Greece}

\date{\small Preprint version — October 2025 \\[4pt]
\textit{This manuscript was completed in June 2025 and first submitted to peer-reviewed venues before being posted as a preprint.}}

\begin{document}

\maketitle

\begin{abstract}
The increasing number of data a booking platform such as Booking.com and AirBnB offers make it challenging for interested parties to browse through the available accommodations and analyze reviews in an efficient way. Efforts have been made from the booking platform providers to utilize recommender systems in an effort to enable the user to filter the results by factors such as stars, amenities, cost but most valuable insights can be provided by the unstructured text-based reviews. Going through these reviews one-by-one requires a substantial amount of time to be devoted while a respectable percentage of the reviews won't provide to the user what they are actually looking for.

This research publication explores how Large Language Models (LLMs) can enhance short rental apartments recommendations by summarizing and mining key insights from user reviews. The web application presented in this paper, named "instaGuide", automates the procedure of isolating the text-based user reviews from a property on the Booking.com platform, synthesizing the summary of the reviews, and enabling the user to query specific aspects of the property in an effort to gain feedback on their personal questions/criteria.

During the development of the instaGuide tool, numerous LLM models were evaluated based on accuracy, cost, and response quality. The results suggest that the LLM-powered summarization reduces significantly the amount of time the users need to devote on their search for the right short rental apartment, improving the overall decision-making procedure.
\end{abstract}

\noindent\rule{\textwidth}{0.4pt}

\section{Introduction}
In early years, looking for a rental property required from the traveler to either arrive at the destination and then roam around asking for suggestions or use the white-pages prior arrival. This procedure was time-consuming, lacked of (objective) reviews, and of sufficient number of choices.

In the past few years, the introduction of hotel booking platforms such as Booking.com and AirBnB has contributed a lot on the tourism sector enabling the tourists to locate a place to spend their nights during their trips in an easy and fast way. Such platforms not only provide a big list of sleeping establishments but they also offer filters and reviews to further augment the user's experience and help them choose the one according to their personal criteria.

According to Eurostat\cite{Eurostat2024}
, in the past few of years, these platforms have attracted an increasing number of users, both from the side of the tourist and from the side of the entity that offers the rental establishment. As a result, the number of establishments being offered on this platform has risen significantly alongside with the total number of reviews each and every establishment has. This makes it time consuming for the interested parties to read enough reviews from each and every one of the rental properties that match the user's criteria applied through the platform's filters such as the star ratings, the presence of private bathroom, etc. At the same time, it is getting more challenging for the establishment's owner as well to keep up with the reviews of their own establishment in an effort to utilize the feedback to further advance the services they offer.

To advance the user's experience in this perspective, this paper introduces a web application called "instaGuide", an AI-powered platform that summarizes unstructured text-based user reviews from a Booking.com establishment. At the same time, it also enables the user to ask specific questions such as "Is the WiFi fast?", "Is it pet friendly?", or "Is free parking available?".

The key contributions of this paper are the development of an LLM-based review summarization system that can also provide an answer in a user's specific question, the evaluation of different LLM models in terms of accuracy, cost, and performance, and the results of feedback received from a small test group. Results reveal that the suggested tool significantly reduces time needed for the user to evaluate each property improving decision-making efficiency.

\section{Related Work}
The "instaGuide" application provides a recommender system\cite{dong2022brief,roy2022systematic,park2012literature,lu2012recommender,alamdari2020systematic,castells2023recommender} that is based on a combination of Data Retrieval and LLM\cite{zhao2023survey,chang2024survey} enhanced with RAG\cite{lewis2020retrieval,cuconasu2024power,gao2023retrieval} technologies. The need to extract meaningful insights from textual-based reviews does not constitute a newly introduced need. Researchers have been putting effort to fulfill this particular need for many years, leading to noteworthy advancements in technology.

\subsection{Recommender Systems Prior to the LLM Era}
Prior to the introduction of the LLM and the RAG technology, literature review on recommender systems based on user-generated reviews in research publications \cite{chen2015recommender,chen2015augmenting,hernandez2019comparative,benabbes2022recommendation,hasan2024reviewbasedrecommendersystemssurvey} explored techniques to mine valuable information from reviews such as collaborative filtering and content-based filtering.

These technologies have been and are still being utilized but they present some shortcomings when rating sparsity and new users/items are introduced in the system. These shortcomings can be overcome by putting effort into extracting information from unstructured text (textual-based reviews) utilizing techniques such as sentiment analysis and features-based recommendations. These techniques rely on opinion mining and natural language processing (NLP)\cite{sun2017review} to advance the recommendations results.

Literature review on sentiment analysis\cite{wankhade2022survey,poria2020beneath} suggests that sentiment analysis face some challenges when it comes to ambiguous situations such as irony, comedy, and sarcasm. For this end, some methods have been introduced\cite{medhat2014sentiment,castro2019towards,poria2018multimodal} but the outcome seems to be not satisfactory enough due to the challenging nature of detecting the aforementioned ambiguous ways of human expression. Other key challenges of sentiment analysis are the informal style of writing and grammatical errors.

Feature-based recommendation systems\cite{han2005feature} are widely used on various platforms such as eshops, on short rental apartment platforms, and generally on platforms that study and rank/compare products/services based on explicit features. For example, the mobile phones comparison platform kimovil.com compares devices with each other by putting side to side their key features (CPU, RAM, screen, etc.). In the case of Booking.com platform, the explicit features consists of the room size, the location, the cleanliness, etc. This method, while it can handle new users and the low number of reviews on a platform, requires manual declaration of the explicit features which are distinct for each sector (e.g. mobile phones, short rental apartments or video games).

A more advanced technique utilizing machine learning is topic modeling\cite{vayansky2020review,churchill2022evolution,abdelrazek2023topic,alghamdi2015survey}. This technique can discover patterns occurring in text and group them into topics (clusters) offering a way to extract structured insights from unstructured text-based user reviews. A topic consists of words that appear frequently in a text and constitute a meaning. For example, in the sector of rental apartments, the topic of "amenities" constitutes of words like pool, spa, gym, WiFi, and parking. Two of the most popular approaches to topic modeling are Latent Dirichlet Allocation (LDA) and Non-Negative Matrix Factorization (NMF). Seemingly, the Amazon marketplace utilizes the topic modeling technique for the products in its marketplace alongside its AI-based summary review generation as depicted in figure 2. In this instance, the topic modeling technique has outlined some patterns on the product's user reviews namely "Functionality", "Ease of setup", "Picture quality", "Solar charging", etc. While this method is quite advanced and offers a good enough result, it doesn't enable the user to ask questions.

\subsection{LLM with RAG technology}
To overcome the need of studying numerous text-based user reviews to reach a safe conclusion while at the same time offering an interactive and personalised user experience, the combination of Large Language Models (LLMs) and Retrieval-Augmented Generation (RAG) has been introduced. Both Amazon and Skroutz.gr marketplaces, utilize the RAG technology to expand the scope of the system's knowledge base to include the reviews and information presented on a product's page.

\begin{figure}[htbp]
  \centering
  \includegraphics[width=\linewidth]{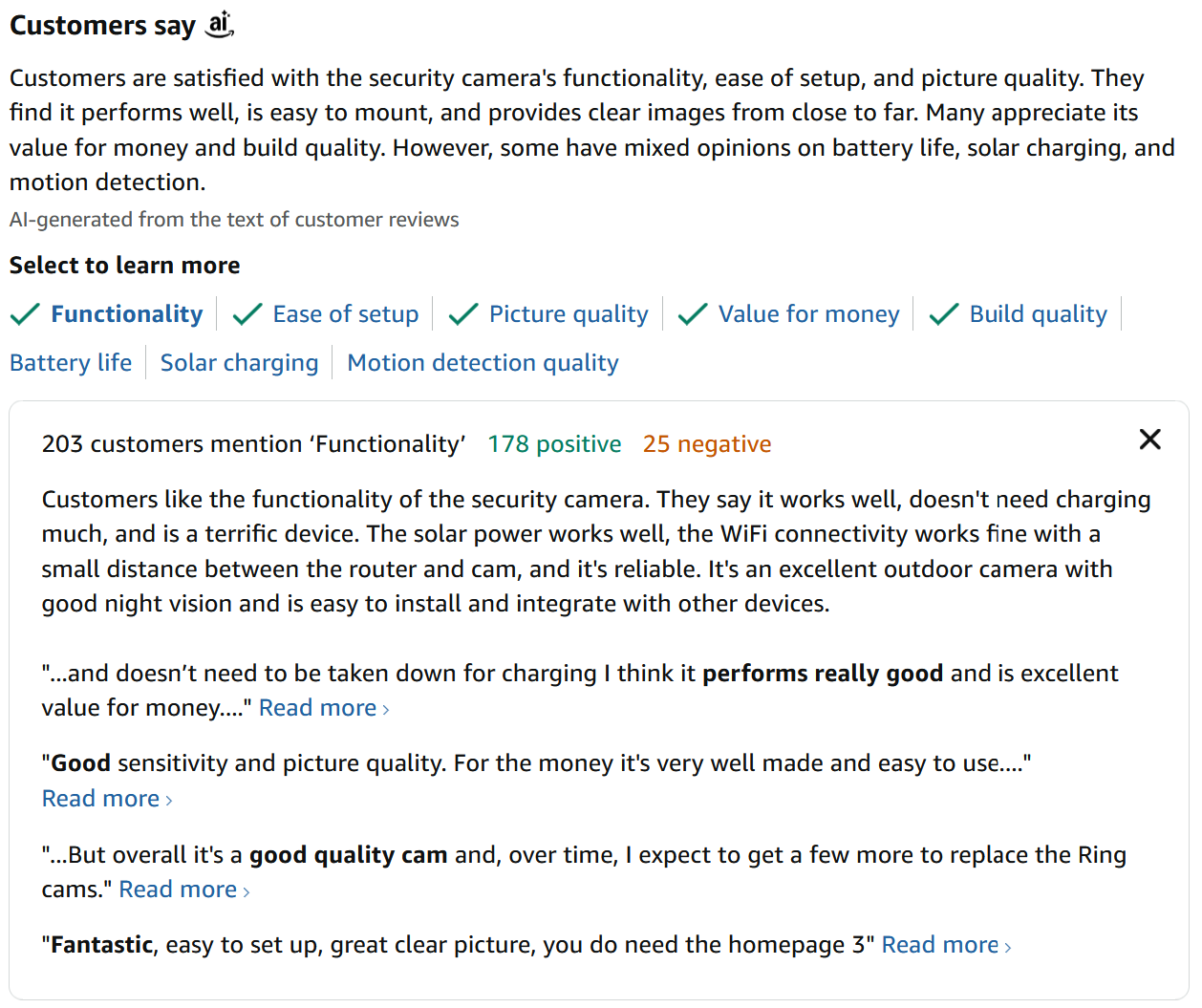}
  \caption{Amazon.co.uk marketplace reviews summary for product eufy Security SoloCam S220 (\url{https://www.amazon.co.uk/eufy-Security-Continuous-Resolution-Compatible/dp/B0BZ4G7S3H/}). Accessed on March 2, 2025.}
  \label{fig:amazon_summary}
\end{figure}

In the case of the Amazon marketplace, a summary of the reviews is generated and presented above the reviews on a product's page accompanied by the title "Customers say" and a combination of the company logo's arrow and the letters "ai". This system, instead of providing a prompt for the user to ask anything related to the product, lists some keywords that enable the potential buyer to find more information on the product in relation to these specific aspects (see Figure~\ref{fig:amazon_summary}). Amazon also announced the launch of its generative AI-powered shopping assistant, named Rufus, in beta version across Europe\cite{amazon-chatbot}. According to Amazon, Rufus can guide the customer on their shopping needs by answering questions related to products. It can make comparisons between products and reach conclusions from these comparisons tailored to the user's needs.

The Skroutz.gr marketplace (\url{https://www.skroutz.gr/}), founded in 2005 in Greece, is the biggest Greek marketplace gaining more than 28 million visits per month and contains more than 30 million products (March 2025)\cite{SkroutzHistory,SkroutzStats}. It follows a slightly different business model than the Amazon marketplace according to which it does not directly sell products rather than provides a search engine to find stores that sell product the user is interested in. For every product, the Skroutz Marketplace lists its description, specifications, reviews, and the list of stores that sell it alongside their price and stores' reviews.

In the Skroutz.gr marketplace, there is no reviews summary automatically generated for each product. Instead, there is a prompt for the platform's user to utilize the AI Assistant (currently in Beta version) by providing two pre-defined questions ("Is it worth it? What users are saying." and "Tell me in a few words what it offers."). Clicking on either of these two questions, the pop-up of the AI Assistant comes up responding in Greek language, providing a summary of the reviews listing both the positive and the negative elements of the product alongside a conclusive remark. The user can also ask any question on the product writing free text. In case the system cannot find enough data from the product's description, specification, or the reviews that can answer the user's query, it communicates the lack of data, prompting the user to review the manufacturer's manual.

If a user were to ask the Skroutz.gr AI Assistant to provide its overview on whether a product, the TP-Link Tapo C320Ws v2.2 for example, is worth it and what the buyers are saying, the AI Assistant responds by saying that the product has received very positive feedback, with a score of 4.75 out of 5 from a total of 373 reviews. According to the buyers' reviews, they appreciate the ease of setup, the video quality both in day and night, and the build quality. The AI Assistant does not limit its overview on the positive elements of the product though. It mentions that most of the negative reviews focus on the camera's sensitivity in movements which raises a notification to the user's mobile device resulting in false positive alarms. At the same time, some users reported random disconnections which seem to be solved with a soft reboot of the device or by replacing the device's power brick and cable. Finally, it provides an overall assessment considering its price, specifications, and the reviews it has received.


The impact of the combination of the LLM with RAG technologies can be observed in various domains. Arslan et al. \cite{arslan2024survey} provide a list of 51 diverse applications of RAG spanning across a wide range of domains such as medical, financial, educational, software development support systems, e-commerce, SQL translation, report generation, and pharmaceutical regulatory compliance inquiries. Similarly, Wenql Fan et al.\cite{fan2024survey} provide a list of RAG enhanced LLM-based applications spanning in the sectors of entertainment, science, and finance. Both of these survey papers highlight how RAG-enhaned LLM applications improve the accuracy and reliability of their outputs by not relying solely on their internal knowledge.

\section{Methodology}
As aforementioned, the suggested recommender system consists of two sub-systems: the one that retrieves the reviews from a listing on Booking.com and the one that summarizes these reviews and can provide answer to the user's queries.

\subsection{Reviews Retrieval}
In an effort to retrieve the customers' reviews of a Booking.com listing, the research team focused on two techniques: a) web scraping, and b) the utilization of a third-party API solution.

\subsubsection{Web Scraping}
The Web Scraping technique\cite{khder2021web,mitchell2018web} has been around for many decades now. Utilizing this technique offers an automated way to fetch information from websites by crawling a website's source code in an effort to locate the HTML element that contains the desired information.

While Web Scraping is being utilized in modern times for several reasons such as products' price monitoring on various e-shops and gathering and consolidating news information\cite{diouf2019web}, this technique's legality and ethics is debatable\cite{krotov2020tutorial}. In the case of the Booking.com platform, its Terms of Service, lastly update in October 31 2023, dictate that no one is allowed to scrape/crawl information from the platform for commercial use without written permission of Booking.com or its licensors\cite{bookingtos}.

\subsubsection{Third Party API solution}
Throughout the tests, the Booking.com Reviews Scraper of both Arel Ventures\cite{arelapi} and of Caprolok\cite{caprolokapi} residing in the Apify Store\cite{apifystore} were utilized. Both tools enable the user to retrieve reviews of short rental listings on the Booking.com platform in a fast and simple manner. Both of these tools provided the functionality necessary for the application's needs.

Both aforementioned APIs can be set up and configured through the Apify API with Python code. To use the Apify API, the user needs to create an account and enable this functionality through their profile's settings. The APIs tested are not free to use but Apify provides 5 USD per month as free balance for every user.

Regarding the Arel Venture's solution\cite{arelapi}, it accepts as arguments a single Booking.com listing, the amount of reviews to fetch, and the desired sorting of review scores (e.g. by date or by review score). The information the API returns in JSON, CSV, XML, Excel or HTML format contains each review's publication date, review score, the feedback's title, positive and negative comment, and the listing manager's reply.

When it comes to Caprolok's solution\cite{caprolokapi}, it is worth noting that this API enables the developer to acquire reviews from multiple listings via a single API call. In addition to what the Arel Venture's API accepts as arguments, this offering accepts multiple Booking.com links of the listings from which it will fetch the reviews from, the range of review scores, the year the reviews were published, and finally, the reviewer's language and type. The data the API call will return for every single review in a JSON, CSV or Excel format contains the hotel's and room's identification number, reviewer's information (username, avatar, country of origin, type), booking's information (number of nights spent, check-in and check-out date), and the review's information (publication date, score, title, positive and negative feedback, the listing's reply on the feedback, number of likes the review acquired, and photographs).

\subsection{Integration of Large Language Models}
Having acquired the listing's reviews, the next step is to summarize them and enable the user to ask custom questions. Large Language Models\cite{zhao2023survey,chang2024survey}, infused with NLP\cite{sun2017review} and RAG technology\cite{lewis2020retrieval,cuconasu2024power,gao2023retrieval}, provide the desired functionality. For the purpose of instaGuide, a number of LLMs were tested.

\subsubsection{Evaluated LLMs}
The first LLM to be tested was OpenAI's GPT-4\cite{gpt4} publicly released on March 2023\cite{gpt4-research}. This model, based on the transformer \cite{amatriain2023transformer,kalyan2021ammus} technology, has been pretrained on predicting the next element (token) on a document, utilizing both publicly available data and data licensed and provided by third-party providers. This model was further optimized (fine-tuned) utilizing the RLHF (Reinforcement Learning from Human Feedback)\cite{kaufmann2023survey} method. While the GPT-4 model provided great advancements on the scene, it still has some shortcomings like hallucinations, limited context window, and lack of "thinking slow" capabilities\cite{achiam2023gpt,mao2023gpteval}.

The next LLM was OpenAI's GPT-4o publicly released on May 2024\cite{gpt4o}. This model, characterized as an auto-regressive omni AI model, introduces advancements over the previous OpenAI's models in terms of image and audio understanding and offers much better interaction in non-English languages. In a more detailed manner, GPT-4o can process text, image, audio, and video, based on a single unified model rather than relying on multiple separate systems working together that handle different types of inputs. As a result, it offers much faster response times while at the same time the cost of using it is significantly reduced. Furthermore, it offers an improved tokenizer, a larger context window and responds with lower latency and higher throughput. A significant advancement of this model is the integration of RAG technology\cite{lewis2020retrieval,cuconasu2024power,gao2023retrieval} which greatly enhances the model's accuracy and relevance by extracting knowledge from external sources. This results in factual consistency, reduced hallucinations, and better handling of domain-specific queries \cite{yue2024mmmu}.

In an effort to make GPT-4o accessible to more people, OpenAI published on July 2024 a minified version of it called GPT-4o mini promising a cost-efficient alternative to the base model. This model provided higher scores in various benchmarks (e.g. MMLU, MGSM, MMMU) when compared to other small models such as Gemini Flash and Claude Haiku\cite{gpt4o-mini}.

On September 2024, OpenAI released the o1-preview\cite{o1-preview} and o1-mini\cite{o1-mini} models. The main difference these models introduced is that they are designed to take more time thinking before they respond to the user's query. This enables the model to analyze and solve problems following a step by step approach making them a good choice for solving complex problems in the fields of math, sciences, and coding. Benchmarked in various cases by OpenAI, the o1-preview model proved to perform similarly to PhD students on demanding physics, chemistry and biology benchmarks. The model was also found to excel in math and programming. Competitions in various fields such as Math, Coding, and in PhD-level science questions (AIME 2024, Codeforces, GPQA Diamond respectively) proved that both models rank much higher than GTP-4o.

Anthropic's Claude 3.5 Sonnet\cite{claude-sonnet}, published on June 2024, was also tested on instaGuide. Anthropic's efforts during the development of this model aimed to offer a model that has the best possible balance between intelligence and cost of use. This model outperformed GPT4o, Gemini 1.5 Pro, and Llama-400b in most of the benchmarks such as GPQA, MMLU, and HumanEval.

On September 2024, Meta published the Llama 3.2 3B model\cite{llama}. The Llama models' main characteristic compared to aforementioned solutions is that they are open source models. The Llama 3.2 3B model is a lightweight choice that only accepts text as input and provides text as output. It can be utilized by external services for automated procedures such as sending event invitations and summarizing text. Being an open source model, it can be executed offline locally on a capable machine ensuring privacy requirements are met.

On May 2024 Google published Gemini 1.5 Flash\cite{gemini}. According to Google's Gemini Team\cite{team2024gemini}, Gemini 1.5 Flash offers a low-latency and high throughput experience offering fast and cost-effective responses to user's queries. Its performance in reasoning accompanied with the context windows of 1.048.576 tokens, is a great choice for interactive applications such as smart agents and real-time translation and interpretation tools.

\subsubsection{Pre-Assessment of LLMs}

\begin{table*}[htbp]
  \centering
  \resizebox{\linewidth}{!}{
      \begin{tabular}{lccccc}
        \toprule
        \textbf{Model} & \textbf{Release Date} & \textbf{Cost (USD per 1M tokens)} & \textbf{Context Window (Prompt / Completion)} & \textbf{Open-Source} \\
        \midrule
        GPT-4 & March 2023 & 30 / 60 & 8,192 / 8,192 & No \\
        GPT-4o & May 2024 & 2.50 / 10 & 128K / 16,384 & No \\
        GPT-4o mini & July 2024 & 0.15 / 0.60 & 128K / 16,384 & No \\
        OpenAI o1-preview & September 2024 & 15 / 60 & 128K / 32,768 & No \\
        OpenAI o1-mini & September 2024 & 3 / 12 & 128K / 65,536 & No \\
        Claude 3.5 Sonnet & June 2024 & 3 / 15 & 200K / 8,192 & No \\
        LLaMA 3.2 3B & September 2024 & Free & 128K / 2,048 & \textbf{Yes} \\
        Gemini 1.5 Flash & May 2024 & Free & 1M / 8,192 & No \\
        \bottomrule
      \end{tabular}
  }
  \caption{Comparison of Key Characteristics of Popular LLMs (as of July 2024)}
  \label{tab:llm-comparison}
\end{table*}

Table~\ref{tab:llm-comparison} presents a comparative overview of the Large Language Models (LLMs) tested during the development of the instaGuide solution. The comparison focuses on three criteria: cost per million tokens, context window size, and open-source availability. These characteristics are important factors in deciding the best fit for tasks such as summarization, dialogue, and query-based information retrieval, particularly in real-time or cost-sensitive environments.

Among the aforementioned models, Gemini 1.5 Flash and GPT-4o mini seem to offer a strong balance of performance, extended context support, and low or no cost, making them particularly attractive for scalable applications like instaGuide where the performance and the cost are crucial. While most models are proprietary and accessed via paid APIs, LLaMA 3.2 is the only open-source model offered, making it the only solution in cases where deploying the LLM locally is crucial in an effort to ensure privacy. The choice of model ultimately depends on the application's specific requirements, such as latency, budget, and privacy concerns.

Among the LLMs listed on table~\ref{tab:llm-comparison} that were tested as a part of instaGuide's deployment, GPT 4o, Claude 3.5 Sonnet, and Gemini 1.5 Flash proved to be the best choices considering that they were the ones to provide the optimal relationship between execution speed and the quality of the returned results. Among the three aforementioned models, Google's Gemini 1.5 Flash proved to be the best choice overall, primarily due to its significantly higher speed.

Finally, the cost of running each one of these models played an important role. According to the tests that took place during the development and evaluation of the instaGuide solution, GPT 4o costs approximately 0,04 euro per response (whether it is to generate the reviews' summary or to reply to a custom query), Claude 3.5 Sonnet costs approximately 0,06 euro whereas Google's Gemini 1.5 Flash is free to use as long as certain conditions are met. The limitations that Google imposes to its Gemini 1.5 Flash API free of cost use consist of a maximum of 15 requests per minute and 1.500 per day and processing no more than 1 million tokens per minute. These limitations do not pose a problem for the needs of developing, testing the capabilities of, and evaluating instaGuide in the academic setting.

\subsection{Development Environment}
The instaGuide application is written on Python utilizing the Django web framework\cite{django}. Python has been proven a great choice for applications that rely on LLM and data processing due to the extensive ecosystem of libraries for Natural Language Processing and Machine Learning. Furthermore, it is a beginner-friendly programming language with a large community of active developers offering rapid prototyping and easy maintenance - particularly important in academic and research settings.

The Django web framework is a well-established framework that provides a ready-to-go environment that offers secure and scalable architecture with support for routing, database modeling, user management, and administrative tools. It also provides easy separation between the front end, back end, and data processing elements of the application's environment.

Taking into consideration the facts listed and expanded on section 3.2 and the prioritization of utilizing a no-cost solution to develop and deploy an application for academic/research purposes, the Web Scraping technique alongside Gemini 1.5 Flash were used.

To ensure the application's portability and consistent operation in multiple environments and to confirm error-free operation in the future regardless of the state of the dependencies' repositories and newer versions, the final project application was containerized utilizing the open source platform Docker\cite{docker}. This container includes all of the required code, libraries, and configuration for the web application to be executed ensuring an isolated and a reproducible environment, independent of dependencies' changes that can result in future conflicts.

\subsection{Review Summarization}

As aforementioned, the generation of the summary of a listing's reviews is accomplished utilizing the RAG technology. This technology enables an LLM to enrich its knowledge base with external material provided by a third party source. In the case of instaGuide, this third party source is the Booking.com providing - willingly enough for academic purpose - its reviews through Web Scraping.

\begin{figure}[H]
  \includegraphics[width=\textwidth]{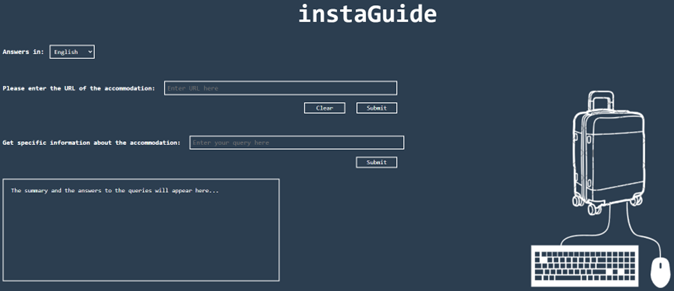}
  \caption{instaGuide web application home page}
  \label{fig:teaser}
\end{figure}

\begin{figure}[htbp]
  \centering
  \includegraphics[width=\linewidth]{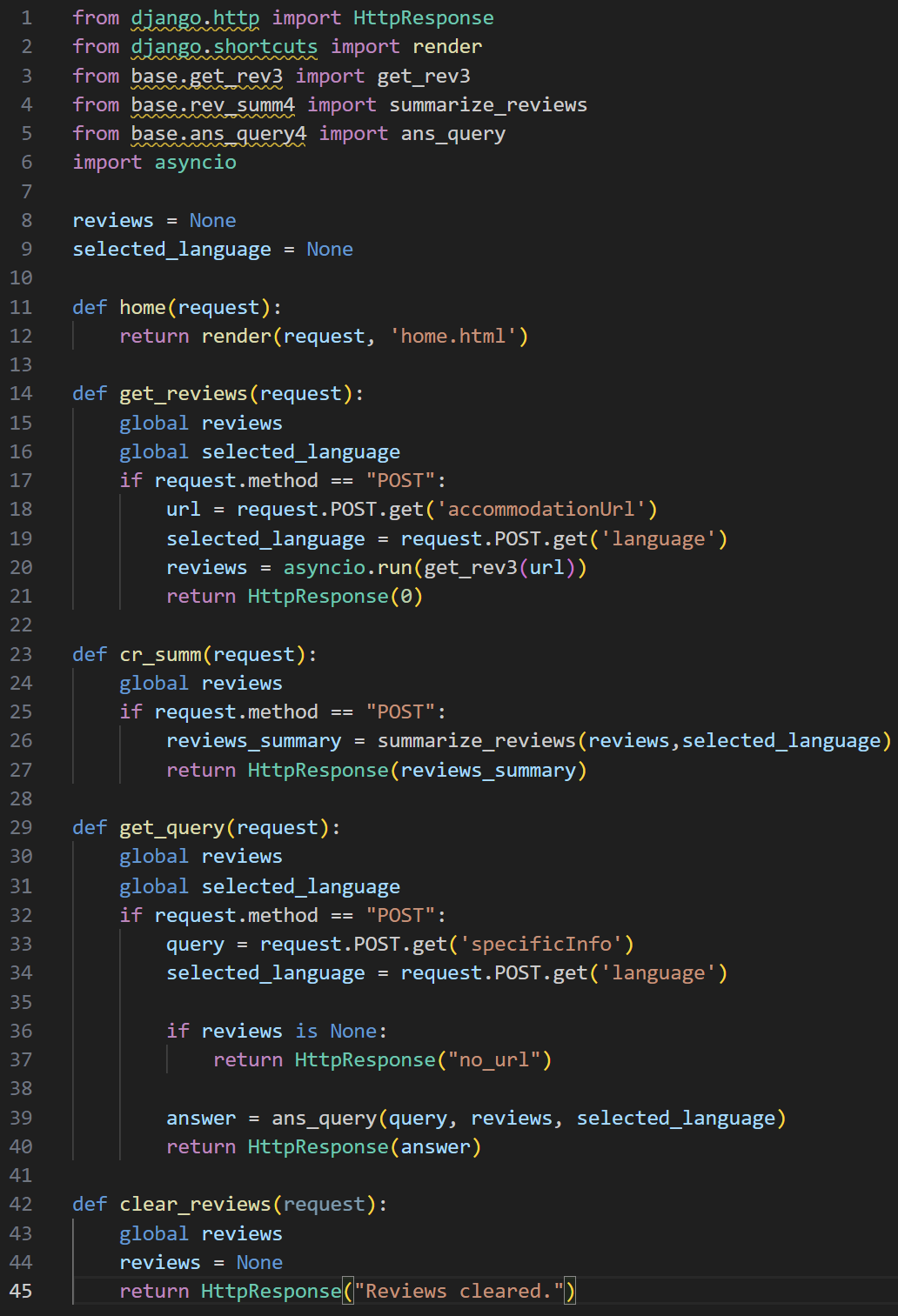}
  \caption{The Python code of the instaGuide's main window seen on Figure 1}
  \label{fig:code-views.py}
\end{figure}

The functionality starts when the users provides a Booking.com listing's url on the appropriate field of Figure ~\ref{fig:teaser} and hits the submit button. At this moment, the instaGuide application fetches the reviews utilizing the \texttt{get\_rev3}
 function as seen on the line 20 of the Figure ~\ref{fig:code-views.py}. Having acquired the accommodation's reviews, the LLM kicks in on line 26 of Figure ~\ref{fig:code-views.py} by summarizing the reviews. The summary of the listing's reviews is presented in the language chosen by the user via the dropdown language switcher placed on the top left part of the instaGuide's main screen as seen on Figure ~\ref{fig:teaser}.

\subsection{Query-Based Retrieval}
\begin{figure*}[htbp]
  \centering
  \includegraphics[width=\textwidth]{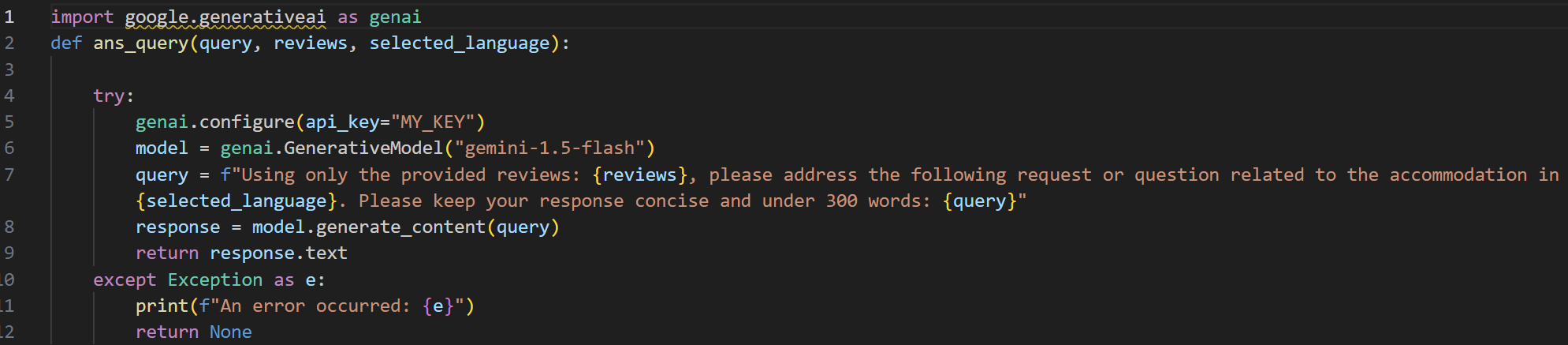}
  \caption{The Python code of instaGuide's query-based retrieval functionality}
  \label{fig:ans_query.py}
\end{figure*}

Having acquired the listing's reviews and generated their summary, the system enables the user to ask any question they want on the listing. This functionality initiates itself on line 29 of Figure ~\ref{fig:code-views.py}. Then, on line 39 of the same code, the function \texttt{ans\_query} is called which takes on the task of providing the fetched reviews, the user's question and the desired language to the LLM and generate the response via careful prompt engineering\cite{clavie2023large,white2023prompt,zhou2022large} as seen on Figure ~\ref{fig:ans_query.py}.

\section{Evaluation}
The application's evaluation was made on a laptop computer with an AMD Ryzen 5 3500U (4 cores, 8 threads, 2.1 to 3.7 GHz)\cite{amd-cpu} and 12GB of RAM. Since the LLMs were deployed on the cloud, the only thing worth mentioning regarding the testing environment is the internet connection speed which was measured at 5.64 Mbps download and 0.74 Mbps upload using Ookla's Speetest\cite{speedtest}. The instaGuide application's source code is available on: *ANONYMIZED REPOSITORY*.

\subsection{Assessment of Review Retrieval Options}
At the time of carrying out this research, having done extended tests with both APIs, the Arel Venture's API takes about 25 seconds to fetch 200 reviews while the Caprolok's API takes about 35 seconds for the same task. This number can vary and depends on the total number of the listing's reviews, the internet connectivity, and the load of the Apify's ecosystem. The cost of the Arel Venture's API tool is 1.50 USD per 1000 reviews fetched while the Caprolok's API cost is 1 USD per 1000 reviews fetched.

On the other hand, the Web Scraping technique does not incur any cost and takes about 5 seconds to fetch 200 reviews. Regarding the legal concerns this technique is burdened with, in the case of instaGuide there is no violation to the Booking.com Terms of Use whatsoever since this application was created strictly for academic research purposes.

Another important aspect of the comparison between the three solutions is the code's maintenance. Since both of the APIs are offered as a Software as a Service, they are accompanied by documentation, technical support and they are updated to comply with Booking.com future platform's updates. On the other hand, the Web Scraping technique requires frequent updates from the instaGuide's development team in an effort to be up to date with the platform's latest updates in the code.

\begin{table*}[htbp]
    \centering
    \resizebox{\linewidth}{!}{
      \begin{tabular}{lccc}
        \toprule
        \textbf{Criterion} & \textbf{Arel Venture API} & \textbf{Caprolok API} & \textbf{Web Scraping} \\
        \midrule
        Retrieval Time (200 reviews) & ~25 seconds & ~35 seconds & ~5 seconds \\
        Cost & \$1.50 per 1000 reviews & \$1.00 per 1000 reviews & Free \\
        Legal Compliance & Compliant (SaaS via Apify) & Compliant (SaaS via Apify) & Grey area – acceptable for academic use \\
        Maintenance Effort & None (SaaS-managed) & None (SaaS-managed) & High – manual updates needed \\
        Support \& Documentation & Available & Available & None \\
        \bottomrule
      \end{tabular}
  }
  \caption{Comparison of Review Retrieval Methods}
  \label{tab:review-retrieval-comparison}
\end{table*}

\subsection{Assessment of LLM Performance}
The purpose of testing all of the LLMs listed on Table~\ref{tab:llm-comparison} on instaGuide was to evaluate their performance in generating a summary and responding to the user's questions, based on knowledge extracted from 200 reviews per listing. In general terms, none of the listed LLM was distinguished for its responses' content. Summarizing a text and extracting knowledge from it is not considered as a difficult task of a SotA LLM model.

As shown in Table ~\ref{tab:llm-tests-results}, the fastest LLM was Google's Gemini 1.5 Flash requiring 3 seconds to provide the summary of the reviews and another 3 seconds to provide the answer to the user's query. The next fastest LLM was GPT-4o mini requiring 5 seconds for the summary and 4 seconds for the answer. Next in line were GPT-4o (7 to 8 seconds for both prompts),  GPT o1-mini (8 to 9 seconds for both prompts), and Claude 3.5 Sonnet (10 seconds for both prompts). The LLM to finish last in line was GPT-4 which required 10 seconds for the summary and 8 seconds for the query - while processing only 70 reviews due to its limited context window of 8,192 tokens. Unfortunately, efforts in utilizing the LlaMA 3.2 3B LLM failed when run both through the Hugging Face platform\cite{hugging-face} and locally, reaching a time-out after 60 seconds of processing.

Regarding the essence of each LLM responses' content, Claude 3.5 Sonnet offers to the point perfectly structured answers, organizing the main points in bullet points. Distinguishing the positive from the negative aspects of a listing, it begins with the positive aspects, moving on to the negative, and finally reaches a conclusion. When the positive aspects of a listing outweigh the negative ones, it will say something like "Although there are some negative comments about *THIS*, *THAT* is commented on by most as a positive element". In case the majority of the comments are negative, it would begin with the positive aspect and then move on to the negatives.

The OpenAI models GPT o1-preview, o1-mini, and 4o mini offer a very similar to the Claude 3.5 Sonnet model answer. The main difference is its structure: it does not use bullet points to list the elements rather than it separates the content into paragraphs. The first paragraph contains the positive aspects, the second one the negative ones, and the third one the conclusion. The GPT-4 model provides the answer in a more condensed way when compared to OpenAI's aforementioned models, offering only a single paragraph.

Finally, GPT-4o and Gemini 1.5 Flash are more in line with OpenAI's GPT-4 model when it comes to the generated reply. The only difference is that in some cases, it can generate two paragraphs of content instead of one.

\begin{table}[htbp]
    \centering
    \resizebox{\linewidth}{!}{
      \centering
      \begin{tabular}{lcc}
        \toprule
        \textbf{LLM} & \textbf{Summary Time (s)} & \textbf{Query Time (s)} \\
        \midrule
        Gemini 1.5 Flash & 3 & 3 \\
        GPT-4o mini & 5 & 4 \\
        GPT-4o & 7--8 & 7--8 \\
        GPT o1-mini & 8--9 & 8--9 \\
        Claude 3.5 Sonnet & 10 & 10 \\
        GPT-4 & 10 & 8 \\
        \bottomrule
      \end{tabular}
  }
  \caption{LLM Performance in instaGuide (Summarization and Query on 200 Reviews)}
  \label{tab:llm-tests-results}
\end{table}

\subsection{User's Feedback}
A test group of 20 people was given access to the instaGuide application to take part in an informal usability study. The test group consisted of people of 20-50 years old looking to book accommodation for their next trip, with at least mediocre knowledge of computer using, interested in traveling abroad. The research team provided minimal directions and guidance on the use of the instaGuide tool in an effort to evaluate the tool's ease of use and intuitiveness. The feedback received from the test group was consistently positive!

Every single person of the test group pointed out that the instaGuide application managed to save them a great deal of valuable time from the procedure of searching for the right fit for a short rental apartment. At the same time, some participants - particularly older users - expressed a concern over the fact that Artificial Intelligence takes on many tasks humans do which can lead a) to many jobs vanishing (e.g. travel agents) and b) to taking decisions or making actions based on likely biased technology (combination of AI and data).

\section{Discussion}
The research done to implement the proposed solution focused both on the three main functionalities of the system, namely review retrieval options, review summarization, and answering the user's questions, and on the user feedback.

Regarding the review retrieval options, the web scraping technique proved to be the best option considering the speed of fetching the reviews (5 second for 200 reviews) and the lack of cost in contrast to utilizing third-party APIs. On the other hand, web scraping is against the Booking.com Term of Services in commercial settings and requires frequent manual code maintenance considering the Booking.com pages' layout changes. Third-party APIs, namely Arel Venture and Caprolok, offered as a SaaS, take upon themselves the ethical considerations and offer a long-term stable solution by introducing a cost and a slower response time in the equation when compared to web scraping.

In depth research was made in testing various LLM models in an effort to identify the best option in terms of cost, speed, and ability to analyze a satisfactory number of reviews. The results indicate that Gemini 1.5 Flash was the fastest LLM among the ones that were tested as shown on Table ~\ref{tab:llm-tests-results}.

\section{Societal and Ethical Impact}
The proposed solution empowers users to make informed decisions when selecting short-term rentals reducing at the same time the cognitive/information overload that surfaces due to the overwhelming amounts of reviews the listings the Booking.com platform has nowadays. At the same time, utilizing this technique, the user can extract knowledge from all the reviews and not just from the ones highlighted by various platforms (e.g. the "Top reviews" method Amazon utilizes for its e-commerce platform).

Considering that the functionality presented on the research paper (review summarization and query-based retrieval), is yet to be offered by platforms such as Booking.com and AirBnB, while some other platforms recently have provided access to certain users on their AI-powered chatbot assistants enabling them to ask questions on products, academic prototypes such as instaGuide can drive industry adoption of more transparent, user-centered recommender technologies.

To deploy this solution in the real world in a commercial/non-academic setting, certain obstacles must be overcome. The first one is the ethical gray area of web scraping which in the case of Booking.com is in direct violation of the service's Terms of Service. In the case of Booking.com, the best solution would be for the Booking.com platform itself to provide the capabilities presented by the instaGuide application to its users. The second obstacle that requires further research and optimization is the biased\cite{ntoutsi2020bias} or hallucinated LLM outputs\cite{liu2024exploring}. Finally, Artificial Intelligence is accompanied with the risk of over-reliance of human on its outputs\cite{spatola2024efficiency,zhai2024effects}. This has been more evident since generative AI was introduced to the masses. Considering the aforementioned, further evaluation and fine-adjustment of the technique is critical before deploying in real-world scenario.

\section{Future Work}
Despite the low number of users participated on the test group and their short-term interaction with instaGuide, it is evident that the proposed solution's offerings were greatly appreciated. At the same time, some areas of advancement were brought into light.

To begin with, having chosen Booking.com as the introductory platform to implement instaGuide, the proposed solution could be very well be applied on other platforms as well such as AirBnB in the industry of short-rental apartments. The proposed solution can be expanded to other fields as well such as car rental aggregators, e-shops, and marketplaces. Finally, it could also be deployed in in-house organization settings to enable stakeholders (e.g. human resources department and/or managers) to extract meaningful insights from (anonymous) feedback the organization's employees submit.

Furthermore, an option of choosing between various LLM models like the ones tested during the implementation of instaGuide would be appreciated by the users. As some of the people on the small test group stated, it would make them feel like they are more in control of the whole procedure. This feedback highlighted the need for embedding Explainable AI\cite{minh2022explainable,ahmed2022artificial,das2020opportunities} on the instaGuide tool in an effort to provide a transparent solution that can justify the reason why every single output (recommendation) is presented on the platform. 

Finally, utilizing the Kubernetes\cite{shah2019building} ecosystem, could offer easy to manage a cloud-deployed application by scaling it on demand, restarting automatically crashed containers, and having an efficient version control system.

\section{Conclusion}
instaGuide is an LLM-powered tool developed to empower users to extract personalized and meaningful insights from unstructured text-based reviews of short-rental apartments listed on the Booking.com platform.

To achieve this goal, a web application built on Python, Django, and Docker was built that integrates multiple LLMs. This paper presents the research done during the development of the proposed solution and the evaluation results of the LLMs tested alongside the feedback received from an informal usability study of 20 people. The impact reported was that the offered solution managed to reduce decision fatigue by offering the users personalized insights based on their own needs.

The development and deployment of instaGuide in the academic setting demonstrates the potential of integrating LLM alongside RAG technologies in various online services such as aggregators and marketplaces. The results indicate that the need to offer a personalized AI-powered assistant that enhances decision-making in environments where high information overload is apparent to every single user is evident.

To support reproducibility and further development of the proposed application, the system,'s source code can be found on the GitHub repository: https://github.com/NikosBelibasakis/InstaGuide.

%
%
\bibliographystyle{plain}
\bibliography{bibfile}

\end{document}